\documentclass[conference]{IEEEtran}

\usepackage{ifpdf}

\usepackage{cite}

\usepackage{graphicx}
\usepackage{epstopdf} 
\ifCLASSINFOpdf
\else
\fi

\usepackage[cmex10]{amsmath}
\usepackage{amsfonts,amssymb}
\usepackage{xcolor}

\usepackage{algorithmic}

\usepackage{array}

\ifCLASSOPTIONcompsoc
\usepackage[caption=false,font=normalsize,labelfont=sf,textfont=sf]{subfig}
\else
\usepackage[caption=false,font=footnotesize]{subfig}
\fi

\usepackage{dblfloatfix}

\usepackage{newtxtext, newtxmath}
\usepackage{url}
\usepackage{arydshln}
\usepackage{enumerate}
\usepackage[mathcal]{euscript}

\newcommand{\vect}[1]{\mathbf{#1}}

\def\beq{\begin{equation}}
\def\eeq{\end{equation}}

\usepackage{fancyhdr}
\fancypagestyle{firstpage}{
	\fancyhf{}
	\fancyhead[C]{Accepted for publication in GC'17 Workshops - LSASLUB}
}

\graphicspath {{figures/}}
\IEEEoverridecommandlockouts

\begin{document}
%
\title{Device Activity and Embedded Information Bit Detection Using AMP in Massive MIMO}
%
%
%

\author{\IEEEauthorblockN{Kamil Senel and Erik~G.~Larsson}
	\IEEEauthorblockA{Dept. of Electrical Engineering, Link\"{o}ping University, Link\"{o}ping, Sweden\\
		Email: \{kamil.senel, erik.g.larsson\}@liu.se}
	\thanks{This work was supported in part by ELLIIT, Swedish Research Council (VR).}}

%
%


\maketitle

\begin{abstract}
Future cellular networks will support a massive number of devices as a result of emerging technologies such as Internet-of-Things and sensor networks. Enhanced by machine type communication (MTC), low-power low-complex devices in the order of billions are projected to receive service from cellular networks. Contrary to traditional networks which are designed to handle human driven traffic, future networks must cope with MTC based systems that exhibit sparse traffic properties, operate with small packets and contain a large number of devices. Such a system requires smarter control signaling schemes for efficient use of system resources. In this work, we consider a grant-free random access cellular network and propose an approach which jointly detects user activity and single information bit per packet. The proposed approach is inspired by the approximate message passing (AMP) and demonstrates a superior performance compared to the original AMP approach. Furthermore, the numerical analysis reveals that the performance of the proposed approach scales with number of devices, which makes it suitable for user detection in cellular networks with massive number of devices.   
\end{abstract}


%
\IEEEpeerreviewmaketitle

\section{Introduction}\label{sec:Introduction}

Future cellular networks will support  machine type communications 
where massive numbers of devices communicate with the network
infrastructure. Applications include Internet of Things (IoT), and especially wireless
connectivity in transportation, smart cities and factories. The number of 
interconnected devices is projected to be in the order of tens of billions by the year $2020$ \cite{mumtaz2017IotMag}.

\begin{figure}[tb]
 	\begin{center}
 		\includegraphics[trim=0cm 0cm 0cm 0cm,clip=true, scale = .7]{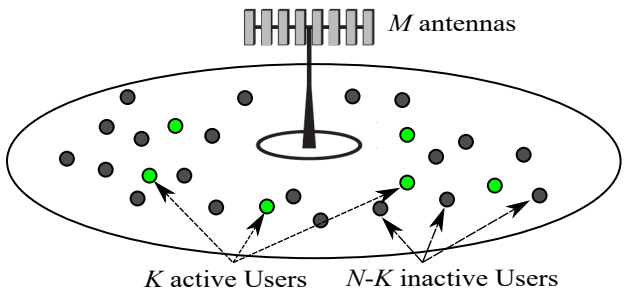}
 		\caption{System setup.}
 		\label{fig:SystemSetup} 
 	\end{center}
 \end{figure}

Systems targeting mainly human-driven traffic have traditionally been optimized for
the transmission of large packets. In contrast, MTC traffic will be dominated by   short
packets \cite{bockelmann2016MTC} (and often, with much lesser requirements on data rates). 
In the MTC context,  the amount of  signaling overhead per packet can  become very
significant compared to the traditional setups with mainly human-driven traffic.
Another key feature of MTC is that it exhibits sporadic traffic patterns, 
i.e., at any given point in time only a small fraction of
the devices are active. One reason for this is 
the inherent intermittency of the traffic (especially for sensor data), but the use
of higher-level protocols that generate bursty traffic also contributes. 
The setup of interest  is depicted in Fig. \ref{fig:SystemSetup}. Here, a base station (BS) with $M$ antennas provides service to $N$ devices and
among these $N$ devices, only $K$  are active at 
a given time. Our focus will be on systems with Massive MIMO technology
such that $M$ is large \cite{redBook}. Massive MIMO
is an important component of the 5G physical layer,
as it enables the multiplexing of many devices in the same time-frequency resources as well as a range extension owing to the coherent beamforming gain.

The intermittency of   MTC traffic calls for efficient mechanisms for random access.
Here we focus on \emph{grant-free random access}, where devices access the network
without a prior scheduling assignment or a grant to transmit. Owing to
the massive number of devices, it is  impossible to assign orthogonal pilot
sequences to every device.
This inevitably leads to collisions between the devices.  Conventionally,
such collisions are handled through  collision resolution mechanisms \cite{lte-book,emil-elisabeth2017RAP}.
A promising class of collision resolution detection methods  
rely on  compressive sensing (CS) for user detection. 

CS based techniques have been proposed as a solution for the joint user activity and data detection problem in sensor networks both under a single antenna \cite{bockelmann2013CS, wang2015CSwNOMA} and MIMO setups \cite{gao2016CSMIMO, garcia2015CSMUD}. However, these works assume the perfect channel information is available and the channel states can reliably be used as a sensing matrix for compressive sensing techniques. A more realistic setup which jointly detects user activity and performs channel estimation is presented in \cite{liu2017Massive}.
These algorithms exploit the sparsity in device activity patterns, and they 
are particularly attractive in  (massive) MIMO setups \cite{gao2016CSMIMO, garcia2015CSMUD,liu2017Massive} thanks to the large number of spatial degrees of freedom in such systems.  

In this work we consider the problem of grant-free random access where the
device has an embedded information bit (EIB), $d$, per block 
to transmit. This bit could be, for example part of  control  
signaling or  an  ACK/NACK bit in a (H)ARQ process (See, e.g., \cite{erik2012piggyALB} for
additional motivation of the scenario). 
Each device is assigned two pilots which are \emph{not} mutually orthogonal, one associated with $d=0$ and
one associated with $d=1$.  In a given slot,
a device is either silent, or it transmits the first pilot (to signal $d=0$) or the second pilot
(for $d=1$). The task of the receiver is to detect active users, estimate their channels and decode the EIB. 

The specific technical contributions of this paper can be summarized as follows: 
\begin{itemize}
	\item We introduce a novel way of embedding an information bit to the pilot sequences to be decoded during the user activity detection process. 
\item We devise a receiver
based on approximate message passing that detects which devices are active, and detects their associated information bit, without using any prior information on neither the channel response nor the user activity.
\item The performance of this AMP-based algorithm is investigated in a massive MIMO setup
($M$ large), and shown to outperform the direct use of an AMP-based user detection technique
\cite{liu2017Massive} followed by independent detection of the information bit, $d$. 
\end{itemize}
Although, throughout this work a single embedded bit is considered exclusively, 
future work will
consider the transmission of multiple (but small numbers of) information bits.

\section{System Setup}\label{sec:SystemSetup}
We consider the uplink communication between a single base station with $M$ antennas and $N$ single antenna users. The channel between user $n$ and the BS is denoted by $\vect{h}_n \in \mathbb{C}^{M\times1}$. The channel is modeled as $\vect{h}_n \triangleq \sqrt{\beta_n}\vect{g}_n$ where $\vect{g}_n$ denotes the Rayleigh fading component and ${\beta_n}$ is the large scale fading. The BS is assumed to know $\beta_n$ as the large scale fading usually varies slowly over time. The transmission from users is assumed to be sporadic with identical activity probability, i.e., in any coherence block each user is active with probability $\epsilon$. 

The channel is assumed to be constant for a duration of $T$ symbols. The active users transmit $L$-length pilots which are utilized for both user detection and channel estimation while the remaining $T-L$ symbols are used for data transmission. In a system with a large number of potential user devices, assigning orthogonal pilot sequences to each user requires $L \geq N$ which may not be feasible owing to the finite channel coherence. This is especially true for future cellular networks with massive number of devices. We assume that the non-orthogonal pilot sequences are generated by sampling an i.i.d. symmetric Bernoulli distribution i.e., the $L$-length pilot sequence of user $n$ is $\vect{s}_n \triangleq [s_{1,n}, \ldots, s_{L,n} ]^T \in \mathbb{C}^{L \times 1}$ where  $s_{l,n} = (\pm 1 \pm j)/\sqrt{2L}$. The total transmission power is assumed to be identical for each device and is denoted by $\rho_{ul}$.

The composite received signal at the BS, $\vect{Y} \in \mathbb{C}^{L\times M}$ is given by 
\begin{equation}\label{eq:SystemModel-1}
\vect{Y} = \sqrt{\rho_{ul}}\sum_{n=1}^N \alpha_n \vect{s}_{n}\vect{h}_{n}^H + \vect{Z}, 
\end{equation}
where $n$ denotes the user index, $\vect{Z} \sim \mathit{CN}(0, \sigma^2 \vect{I})$ is the additive white Gaussian noise; $\alpha_n$ is the user activity indicator with $\text{Pr}(\alpha_n = 1) = \epsilon$ and $\text{Pr}(\alpha_n = 0) = 1 - \epsilon$. Among $N$ users only $K$ are active in a given coherence block. Let 
\begin{equation}
\vect{x}_n = \alpha_n \vect{h}_n, \quad \forall n = 1, \ldots, N.
\end{equation}
 Rewriting \eqref{eq:SystemModel-1}, we obtain
\begin{equation} \label{eq:SystemModel-2}
\vect{Y} = \sqrt{\rho_{ul}}\vect{S}\vect{X} + \vect{Z}
\end{equation}  
where $\vect{S} = [\vect{s}_1, \ldots, \vect{s}_N]$, $\vect{X} = [\vect{x}_1, \ldots, \vect{x}_N]^H$ and
 
Note that if user $n$ is inactive $\alpha_n = 0$ and the corresponding row $n$ of $\vect{X}$ is zero which results in a sparse structure as $\vect{X}$ has $\epsilon N$ non-zero rows on average.
   

The motivation of this work is based on finding efficient ways of physical layer control signaling and grant-free random access with small amounts of data in mobile systems. Specifically in control signaling it is often of interest to send single control bits \cite{erik2012piggyALB}. 
%
  Specifically we propose an approach which assigns multiple pilot sequences which are utilized to transmit EIBs during the user detection and channel estimation process.  


\section{Review of Approximate Message Passing}
The problem of detecting and estimating the non-zero rows of $\vect{X}$ based on the noisy observations, $\vect{Y}$ and known pilot sequences is a compressive sensing problem. For the single antenna setup \eqref{eq:SystemModel-2} the problem reduces to the single measurement vector (SMV) reconstruction problem whereas with multiple antennas it becomes a multiple measurement vector (MMV) reconstruction problem. In this work, we utilize an algorithm with low complexity called approximate message passing \cite{jongmin2011Beliefprop} to recover the sparse $\vect{X}$. Next, we briefly review the AMP algorithm.

 Let $t$ denote the index of the iterations and $\hat{\vect{X}^t} = [\hat{\vect{x}}^t_1, \ldots, \hat{\vect{x}}^t_N]^H$ be the estimate of $\vect{X}$ at iteration $t$. Then, the AMP algorithm can be described as follows
\begin{eqnarray}
\hat{\vect{x}}^{t+1}_n &=& \eta_{t,n} \left( (\vect{R}^t)^H \vect{s}_n + \hat{\vect{x}}^{t}_n \right) \label{eq:AMP-1}\\
\vect{R}^{t+1} &=& \vect{Y} - \vect{S}\hat{\vect{X}}^{t+1} + \frac{N}{L} \vect{R}^t \sum_{n=1}^N \frac{\eta_{t,n}' \left( (\vect{R}^t)^H \vect{s}_n + \hat{\vect{x}}^{t}_n \right)}{N} \label{eq:AMP-2}
\end{eqnarray}
where $\eta(.)$ is a denoising function, $\eta(.)'$ is the first order derivative of $\eta(.)$ and $\vect{R}^t$ is the residual at iteration $t$ \cite{rangan2016vectorAMP}. An important advantage of AMP is that in the asymptotic region, i.e., as $L,~K,~N \rightarrow \infty$, the behavior is described by a set of state evolution equations \cite{rangan2011GeneralziedAMP}. In vector form, the state evolution is given by \cite{bayati2011dynamics}
\begin{equation}\label{eq:sigmaUpdate}
\Sigma^{t+1} = \frac{\sigma^2}{\rho_{ul}}\vect{I} + \frac{N}{L} \mathbb{E}\{ \|\eta (\vect{x}_{\beta} - (\Sigma^t)^{\frac{1}{2}}\vect{w})-\vect{x}_{\beta} \|^2\}
\end{equation}    
where $\vect{w}\in \mathbb{C}^{M\times1}$ is a complex Gaussian vector with unit variance and  $\vect{x}_{\beta} \in \mathbb{C}^{M\times1}$ has the distribution 
\beq
p_{\vect{x}_{\beta}} = (1 - \epsilon)\delta + \epsilon p_{\vect{h}_{\beta}}.
\eeq
Here, $p_{\vect{h}_{\beta}} \sim \mathit{CN}(0, \beta \vect{I})$ is the distribution of the channel vector of the active device and $\delta$ is the dirac Delta at zero corresponding to the inactive device channel distribution. The expectation in \eqref{eq:sigmaUpdate} is taken with respect to $\beta$ and allows the analytical performance analysis of the AMP algorithm as the update given by equations \eqref{eq:AMP-1}-\eqref{eq:AMP-2} are statistically equivalent to applying a denoiser to the following \cite{rangan2011GeneralziedAMP}
\beq \label{eq:equaivalentAMP}
\hat{\vect{x}}^t_n = \vect{x}_n + (\Sigma^t)^{\frac{1}{2}}\vect{w} = \alpha_n \vect{h}_n +  (\Sigma^t)^{\frac{1}{2}}\vect{w},
\eeq       
which decouples the estimation process for different users.
 The state evolution is shown to be valid for a wide range of Lipschitz continuous functions \cite{bayati2011dynamics}. For the multiuser detection problem, the following denoising function is used: 
\begin{equation}\label{eq:etaMMSE}
\eta(\hat{\vect{x}}_n) = t(\hat{\vect{x}}_n^t; \vect{\Sigma}^t )\beta_n\left(\beta_n \vect{I} + \vect{\Sigma}^t\right)^{-1} \hat{\vect{x}}_n 
\end{equation}   
where 
\begin{eqnarray}
t(\hat{\vect{x}}; \vect{\Sigma} ) &=& \frac{1}{1+\frac{1-\epsilon}{\epsilon}\text{det}(\vect{I} + \beta_n (\vect{\Sigma})^{-1})^{1/2} q(\hat{\vect{x}}; \vect{\Sigma} )}, \\
 q(\hat{\vect{x}}; \vect{\Sigma} ) &=& \text{exp}\left(-\frac{1}{2} \hat{\vect{x}}^H(\vect{\Sigma}^{-1}- (\vect{\Sigma} + \beta_n\vect{I})^{-1} )\hat{\vect{x}} \right).
\end{eqnarray}
The denoising function \eqref{eq:etaMMSE} is shown to be the MMSE for the equivalent system described by \eqref{eq:equaivalentAMP} in \cite{jongmin2011Beliefprop}. 
 
Note that $t(\cdot)$ is a thresholding function based on the likelihood ratio which can be computed by considering two cases  \eqref{eq:equaivalentAMP}, user $n$ is active, i.e., $\alpha_n = 1$ and $\alpha_n = 0$ when the user is inactive. In the absence of the thresholding term, \eqref{eq:etaMMSE} reduces to the linear MMSE estimator for the case when every user is active, i.e., $\epsilon = 1$.

\subsection{Review of AMP for User Detection Activity}
The AMP approach heavily relies on the sparsity in the device activity pattern. The so-called ''sparsity-undersampling tradeoff'' states that as sparsity decreases, the length of the pilot sequences must increase in order to achieve the same performance \cite{donoho2009mpa}. For the noiseless case, a lower bound on the length of pilot sequences for perfect recovery is given by $L \geq K$ \cite{jongmin2011Beliefprop,baron2005dcs}. However, for the user detection problem perfect recovery is not essential and there are also works which demonstrates superior performance in terms of user detection for the cases with $L < K$ \cite{liu2017Massive}.  

\begin{figure}[thb]
	\begin{center}
		\includegraphics[trim=.3cm 0cm 0cm 0.4cm,clip=true,width = 9cm]{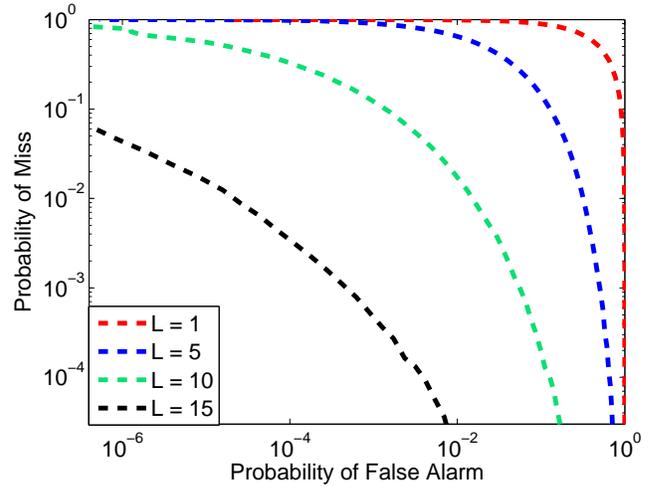}
		\caption{Probabilities of miss and false alarm with respect to pilot sequence length under a setup with $M = 20$, $N = 100$ and $\epsilon = 0.05$.}
		\label{fig:fig2}
	\end{center}
\end{figure}
 
 Fig. \ref{fig:fig2} demonstrates the performance of the AMP algorithm for various pilot sequence lengths under a setup with $M = 20$, $N = 100$ and $\epsilon = 0.05$. The results illustrate the high dependence of the performance on the pilot sequence length. Note that the improvement is especially significant when $L$ is equal to the expected number of active devices and for longer sequence lengths.
 
\begin{figure}[thb]
	\begin{center}
		\includegraphics[trim=.3cm 0cm 0cm 0.4cm,clip=true,width = 9cm]{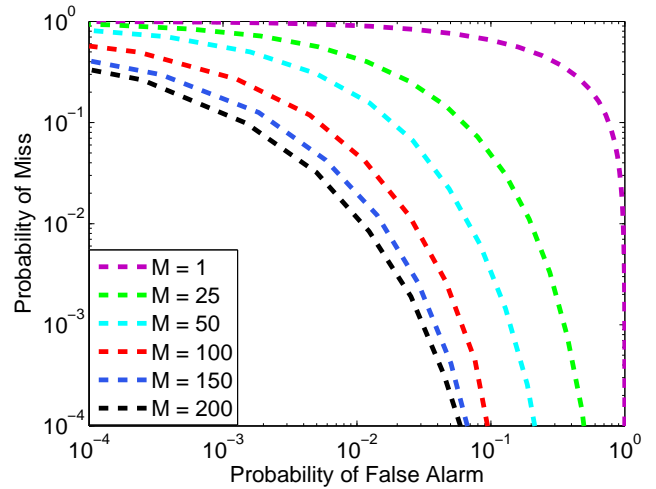}
		\caption{Probabilities of miss and false alarm with respect to number of antennas for $N = 100$ and $\epsilon = 0.05$ with $L = 5$.}
		\label{fig:fig1}
	\end{center}
\end{figure}

 Another crucial parameter which effects the user detection performance, is the number of antennas at the BS. User detection performance of the AMP algorithm with respect to various number of BS antennas is illustrated in Fig. \ref{fig:fig1}. Increasing the number of antennas significantly improves the performance. However, the performance gains due to increased number of antennas experiences a saturation effect, i.e., the improvement gradually decreases as $M$ increases. This shows that increasing number of antennas enhances the performance of AMP algorithm for user detection, however the number of antennas should not be considered as an absolute substitute for pilot sequence length.
 
 
 
%
%
   

\section{New Transmission Scheme and AMP algorithm for EIB Transmission}
The underlying idea of the proposed approach is to convey the EIB while transmitting pilot signals. In order to transmit the EIB, each user is assigned two different pilot sequences and chooses one of them based on the EIB. Let $d_n$ denote the EIB of user $n$; $d_n \in \{0,1\}$. Since the pilot sequences are not orthogonal and are generated from i.i.d. Bernoulli distributions, the effect of transmitting the EIB is manifested through $\vect{S}$. In particular, with EIB transmission equation \eqref{eq:SystemModel-2} becomes
\beq\label{eq:SystemModelALB-1} 
  \vect{Y} = \sqrt{\rho_{ul}}\tilde{\vect{S}}\vect{\tilde{X}} + \vect{Z}
  \eeq
  where $\tilde{\vect{S}} = [\vect{s}_{1,0}, \vect{s}_{1,1}, \ldots, \vect{s}_{N,1}] \in \mathbb{C}^{L\times 2N}$ and $\vect{\tilde{X}} = [(1-d_1)\vect{x}_1, d_1\vect{x}_1,  \ldots, (1- d_N)\vect{x}_N, d_N\vect{x}_N]^H \in \mathbb{C}^{2N\times M}$. Hence, EIB transmission is similar to the previous example illustrated in Fig. \ref{fig:fig2} as number of potential users is increased to $2N$ while the number of active users is unchanged. Since the number of active users is not affected by EIB transmission, the BS needs to detect $K$ users among $2N$ potential users, i.e., $\text{Pr}(\alpha_n = 1) = \epsilon/2$ and $\text{Pr}(\alpha_n = 0) = 1 - \epsilon/2$ for the case of EIB transmission. 
  
  In principle, AMP algorithm could be utilized to detect users along with the EIB without any additional effort. However, such an approach would be strictly suboptimal as it does not utilize the structure of $\tilde{\vect{X}}$. In particular, it is not possible for a user to transmit $2$ pilot sequences concurrently, hence it is known a priori that it is not possible to have the nonzero rows of $\tilde{\vect{X}}$ corresponding to the same user simultaneously. In its original form, described by \eqref{eq:AMP-1}-\eqref{eq:AMP-2},  the AMP algorithm does not utilize this information.   
  
  \subsection{Algorithm Description} 
  
\begin{figure}[thb]
  	\begin{center}
  		\includegraphics[trim=.8cm 0cm 0cm 0.5cm,clip=true,width = 9cm]{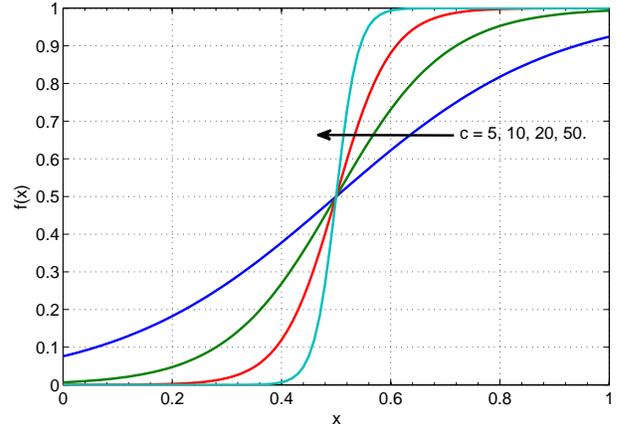}
  		\caption{Sigmoid function for various values of $c$.}
  		\label{fig:fig-sigmoid}
  	\end{center}
  \end{figure}

  In this subsection, the details of modifying the AMP algorithm  in a way to exploit the structural properties of $\tilde{\vect{X}}$, is presented.  Assume that user $k$ is active, i.e., $\alpha_k = 1$, then
  
\begin{flalign} \label{eq:2casesX} 
  \hat{\vect{x}}^t_k = \begin{cases} 
  	 \vect{x}_k + (\Sigma^t)^{\frac{1}{2}}\vect{w}\sim \mathit{CN}(0, \beta_k \vect{I}+ \vect{\Sigma}^t), & \text{if}~~ d_k = 0, \\
  	(\Sigma^t)^{\frac{1}{2}}\vect{w}\sim \mathit{CN}(0, \vect{\Sigma}^t), & \text{if}~~ d_k = 1,
  \end{cases}
  \end{flalign}
  and vice versa for $\hat{\vect{x}}^t_{k+1}$. Notice that, when user $k$ is active depending on $d_k$ either $\vect{x}_k$ or $\vect{x}_{k+1}$ is non-zero. The likelihood function based on \eqref{eq:2casesX} is given by
  \begin{equation}
 \Lambda (\hat{\vect{x}}^t_k) = \frac{|\Sigma^t|^{1/2}}{|\beta_k \vect{I}+\Sigma^t|^{1/2}} q(\hat{\vect{x}}; \vect{\Sigma}^t )^{-1}.
  \end{equation}
  Let $\varphi(\hat{\vect{x}}^t_k)$ denote the EIB coefficient defined by 
  \begin{equation}
  \varphi(\hat{\vect{x}}^t_k) = \frac{\Lambda (\hat{\vect{x}}^t_k)}{\Lambda (\hat{\vect{x}}^t_k) + \Lambda (\hat{\vect{x}}^t_{k+1})}
  \end{equation}
  which can be thought of as a measure of the proportional likelihood of the EIB. The EIB coefficient provides a form of proportional thresholding, however in order to enhance its effectiveness a sharper threshold is required. In the ideal case, the receiver should only decide on one of the two possible EIB sequences while suppressing the other one and in order to achieve this, we utilize a soft-thresholding function known as a sigmoid function. The sigmoid function is defined by 
  \begin{equation}\label{eq:sigmoid}
  f(x) = \frac{1}{1 + \exp(-c(x - \frac{1}{2}))}.
  \end{equation}   
   Here, $c$ allows us to control the sharpness of the sigmoid function which is depicted in Fig. \ref{fig:fig-sigmoid}. The resulting modified denoiser is 
  \begin{equation}\label{eq:modifiedEta}
  \tilde{\eta}(\hat{\vect{x}}_n) = f(\varphi(\hat{\vect{x}}^t_k))t(\hat{\vect{x}}_k^t; \vect{\Sigma}^t )\beta_n\left(\beta_n \vect{I} + \vect{\Sigma}^t\right)^{-1}.  
  \end{equation}
 Note that the modified denoiser is Lipschitz-continuous which is required for the validity of state evolution \cite{bayati2011dynamics}. The proposed modified AMP algorithm 
 (M-AMP) is specifically designed for EIB transmission and utilizes the denoiser given in \eqref{eq:modifiedEta}.
 The principal idea for the modified AMP algorithm (M-AMP) is that EIB transmission via using multiple pilots increases the sparsity of the system in a structured manner. In other words, it is impossible for a user to transmit both pilot sequences at the same time.  
    
  \subsection{Theoretical Analysis}
  The state evolution allows us to analyze the error probabilities with respect to number of antennas $M$ in the asymptotic region. Recall that as $L,~K,~N \rightarrow \infty$ at fixed ratios, the behavior of the AMP algorithm is described by a set of state evolution equations and this property allows us to accomplish the succeeding analysis. First, note that under the uncorrelated channel assumption, we have $\Sigma^t = \tau_t^2 \vect{I}$. The initial value of the state is 
  \begin{equation}\label{eq:InitialState}
  \Sigma^{0} = \frac{\sigma^2}{\rho_{ul}}\vect{I} + \frac{N}{L} \mathbb{E}\{ \vect{x}_{\beta}\vect{x}_{\beta}^H\}
  \end{equation}     
  and for the uncorrelated channels $\Sigma$ is a diagonal matrix with identical elements \cite{liu2017Massive}. This allows us to simplify the AMP algorithm as follows
  \begin{equation}\label{eq:etaMMSEsimplified}
  \eta(\hat{\vect{x}}_n) = t(\hat{\vect{x}}_n^t; \vect{\Sigma}^t )\frac{\beta_n}{\beta_n  + \tau_t^2}  \hat{\vect{x}}_n
  \end{equation}   
  and
  \begin{eqnarray}
  t(\hat{\vect{x}}; \vect{\Sigma} ) &=& \frac{1}{1+\frac{1-\epsilon}{\epsilon}(\frac{\tau_t^2 + \beta_n}{\tau_t^2})^{M/2} q(\hat{\vect{x}}; \vect{\Sigma} )}, \\
  q(\hat{\vect{x}}; \vect{\Sigma} ) &=& \text{exp}\left(-\frac{1}{2} \hat{\vect{x}}^H\hat{\vect{x}}\left(\frac{1}{\tau_t^2} - \frac{1}{\tau_t^2 + \beta_n}\right) \right).
  \end{eqnarray}
  Furthermore, \eqref{eq:equaivalentAMP} becomes 
  \begin{equation}\label{eq:equaivalentAMPsimplified}
  \hat{\vect{x}}^t_n = \vect{x}_n + \tau_t\vect{w} = \alpha_n \vect{h}_n +  \tau_t\vect{w}
  \end{equation}
  which will be utilized when designing the user activity detector. Without loss of generality, consider user $n$ and assume that $d_n = 0$. Similar to the case presented in \eqref{eq:2casesX}, $\hat{\vect{x}}^t_n$ is a complex Gaussian random vector independent of $\alpha_n$. Furthermore, under the assumption that the channels to different antennas can be represented by uncorrelated and identical random variables, $\hat{\vect{x}}^t_n$ has i.i.d. entries. For the case where $\alpha_n = 1$, $\hat{\vect{x}}^t_n$ has variance $\beta_n + \tau_t^2$ and $\tau_t^2$ when $\alpha_n = 0$. Next, consider the likelihood function 
  \begin{eqnarray}\label{eq:likeNoALB}
  \Lambda (\hat{\vect{x}}^t_k) = \frac{P(\hat{\vect{x}}^t_k|\alpha_n = 1)}{P(\hat{\vect{x}}^t_k|\alpha_n = 0)} &\lessgtr& 1\nonumber \\
    e^{- \frac{1}{2}\left(\frac{1}{\beta_n + \tau_t^2} - \frac{1}{\tau_t^2}\right)\|\hat{\vect{x}}^t_k\|^2}&\lessgtr& \left(\frac{\tau_t^2 + \beta_n}{\tau_t^2}\right)^{\frac{M}{2}} \nonumber 
    \end{eqnarray}
    which can be further simplified to obtain
    \begin{equation}\label{eq:threshold}
     \|\hat{\vect{x}}^t_k\|^2 \lessgtr M \ln\left(\frac{\tau_t^2 + \beta_n}{\tau_t^2}\right) \frac{\tau_t^2 \left(\beta_n + \tau_t^2\right)}{\beta_n}.
    \end{equation}
    Note that the threshold given in \eqref{eq:threshold} assumes equal costs to miss and false detection. It is possible to utilize a lower threshold to obtain a lower miss detection probability and vice versa.    
    In the asymptotic region, $\|\hat{\vect{x}}^t_k\|^2$ is $\chi^2$ distributed with 2$M$ DoF and with the threshold given in \eqref{eq:threshold}, it can be shown that both probability of false alarm and probability of miss detection vanishes as $M$ increases \cite{liu2017Massive}.
         
%

\section{Numerical Results} 
In this section, numerical performance analyses of the modified and original AMP algorithm are provided. A single centralized BS with a $350$m cell radius and uniformly distributed users are considered. The path loss model is given as $\beta_k = -130 - 37.6\log_{10}\left( r_k\right)\,$dB where $r_k$ denotes the distance of user $k$ to the BS in kilometers. The transmission powers of the users are assumed to be identical and equal to $\rho_{ul} = 10$dBm. All simulations are carried out with $10$dB SNR and the number of iterations is $20$ for all algorithms. For M-AMP the sigmoid function is realized with $c = 10$ for all simulations.    

\begin{figure}[thb]
	\begin{center}
		\includegraphics[trim=0.5cm 0cm 0cm 0.3cm,clip=true,width = 9.5cm]{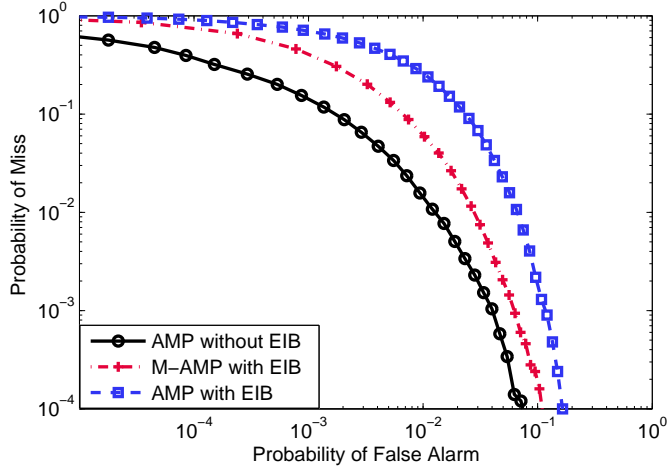}
		\caption{Probabilities of miss and false alarm of user detection for various number of antennas under a setup with $N = 100$, $L = 5$ and $\epsilon = 0.05$.}
		\label{fig:fig3}
	\end{center}
\end{figure}
In Fig. \ref{fig:fig3}, the user detection performance of $3$ different AMP algorithms are depicted. The algorithms compared are as follows:
\begin{itemize}
	\item \textbf{AMP without EIB}: The original AMP algorithm which considers $N = 100$ pilot sequences without any EIB.
	\item \textbf{M-AMP with EIB}: The modified AMP algorithm which considers $N=200$ pilot sequences and detects users along with EIB.
	\item \textbf{AMP with EIB}: The original AMP algorithm which considers $N=200$ pilot sequences and detects users along with EIB.
\end{itemize}
 There are $100$ potential users and on the average only $\epsilon N$ are active. For the case when EIB is transmitted the detector must detect the active users by determining among $200$ pilot sequences. In this case, if the detector determines that one of the pilot sequences corresponding to a user is transmitted, then that user is detected as an active user independently of whether a EIB is transmitted. In all cases, the number of iterations and pilot sequence length are identical. As expected the AMP algorithm without EIB provides the best performance whereas the M-AMP outperforms AMP with EIB. 
  
\begin{figure}[thb]
	\begin{center}
		\includegraphics[trim=0.5cm 0cm 0cm 0.3cm,clip=true,width = 9.5cm]{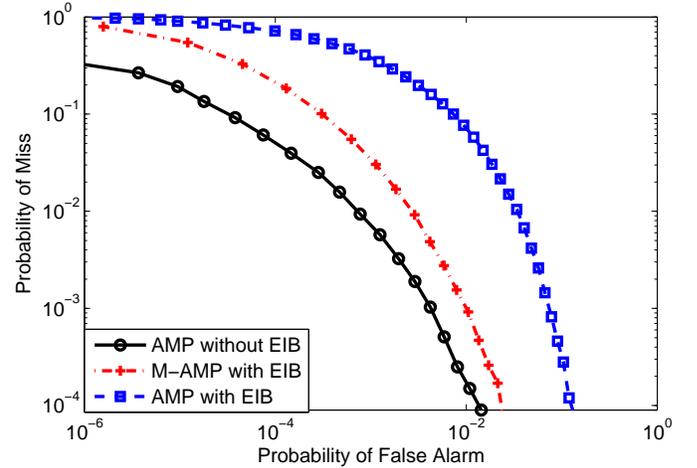}
		\caption{Probabilities of miss and false alarm of user detection for various number of antennas under a setup with $N = 200$, $L = 10$ and $\epsilon = 0.05$.}
		\label{fig:fig4}
	\end{center}
\end{figure}

An interesting property of the AMP algorithm is that increasing $N$, $L$ and $K$ while keeping their ratios fixed improves the performance. In Fig. \ref{fig:fig4} another example is given. The behavior of each algorithm is similar, however the performances of all of the approaches are superior compared to the case with $100$ users. This shows that the proposed approach is scalable for large numbers of users. 
 
\begin{figure}[thb]
	\begin{center}
		\includegraphics[trim=0cm 0cm 0cm 0.5cm,clip=true,width = 9.5cm]{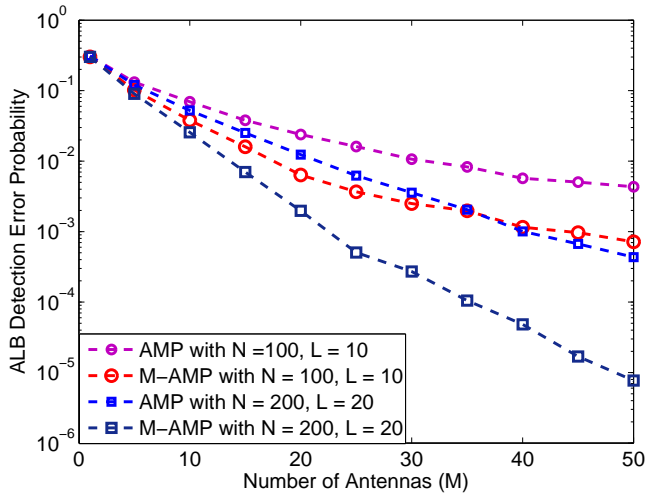}
		\caption{The EIB detection error probabilities as a function of the number of antennas.}
		\label{fig:fig5}
	\end{center}
\end{figure}

Fig. \ref{fig:fig5} demonstrates the EIB detection performance of the AMP and M-AMP algorithms for different number of antennas. In this example, two different setups are considered, the first one with $N = 100$, $\epsilon N = 5$ with $L = 10$ and the second setup with $N = 200$, $\epsilon N = 10$ with $L = 20$.
 The EIB detection is carried out with a threshold that provides equal probability of false alarm and miss detection. In both cases, the M-AMP algorithm outperforms the original algorithm in terms of overall performance and the difference becomes more significant as the number of devices increases. Another observation is that for the single antenna case both algorithms achieve almost identical results and both of the algorithms benefit from an increased number of users which shows that the proposed approach is scalable for a setup with massive number of devices.

\section{Conclusion}
We considered grant-free random access in the form of combined   
user-detection and non-coherent transmission of an embedded information bit per
packet, using a large array of antennas at the receiver.  We proposed
an algorithm based on AMP \cite{jongmin2011Beliefprop}, and showed that it 
outperforms the direct use of AMP-based user detection \cite{liu2017Massive} followed by decoding of the embedded
information bit. Future work should extend our scheme to multiple (but small numbers of)
information bits,
targeting the vision of fully non-coherent communication for MTC in Massive MIMO.

%
%


\ifCLASSOPTIONcaptionsoff
  \newpage
\fi

%

%
%
%




\end{document}